\documentclass{article}
\usepackage{spconf,amsmath,graphicx, hyperref,xcolor,colortbl,subfig, comment, multirow}
\hypersetup{
    colorlinks=false,
    linkcolor=blue,
    filecolor=magenta,      
    urlcolor=cyan,
}

\title{DNSMOS P.835: A Non-Intrusive Perceptual Objective Speech Quality Metric to Evaluate Noise Suppressors}
\name{Chandan K A Reddy, Vishak Gopal, Ross Cutler}
\address{Microsoft Corporation, Redmond, WA\\
    chandan.ka@outlook.com, vishak.gopal@microsoft.com, ross.cutler@microsoft.com}
\begin{document}
%
\maketitle
\begin{abstract}
Human subjective evaluation is the ``gold standard'' to evaluate speech quality optimized for human perception. Perceptual objective metrics serve as a proxy for subjective scores. We have recently developed a non-intrusive speech quality metric called Deep Noise Suppression Mean Opinion Score (DNSMOS) using the scores from ITU-T Rec. P.808 \cite{ITU-P808} subjective evaluation. The P.808 scores reflect the overall quality of the audio clip. ITU-T Rec. P.835 \cite{ITU-P835} subjective evaluation framework gives the standalone quality scores of speech and background noise in addition to the overall quality. In this work, we train an objective metric based on P.835 human ratings that output 3 scores: i) speech quality (SIG), ii) background noise quality (BAK), and iii) the overall quality (OVRL) of the audio. The developed metric is highly correlated with human ratings, with a Pearson's Correlation Coefficient (PCC)=0.94 for SIG and PCC=0.98 for BAK and OVRL. This is the first non-intrusive P.835 predictor we are aware of. DNSMOS P.835 is made publicly available as an Azure service.     
\end{abstract}
\section{Introduction}
\label{sec:intro}

Subjective evaluation of speech quality is the most reliable way to evaluate Speech Enhancement (SE) methods \cite{reddy2019scalable}. However, subjective tests are not easily scalable as they require a considerable number of listeners, the process is laborious, time-consuming, and expensive. Conventional objective speech quality metrics such as Perceptual Evaluation of Speech Quality (PESQ) \cite{941023}, Perceptual Objective Listening Quality Analysis (POLQA) \cite{polqa}, VisQOL \cite{hines_visqol_2015} and Signal to Distortion Ratio (SDR) are widely used to evaluate Speech Enhancement (SE) algorithms optimized for human perception. Some of these metrics are designed to predict the subjective Mean Opinion Score (MOS) obtained using the Telecommunication Standardization Sector of the International Telecommunication Union (ITU-T) Recommendation P.800 \cite{p800}. However, they are shown to correlate poorly with human rating when used for SE tasks that involve perceptually invariant transformations \cite{reddy2019scalable}. Also, intrusive metrics cannot be used to evaluate real recordings when a clean reference is unavailable in realistic scenarios.

\section{Related work}
The subjective test ITU-T P.835 \cite{ITU-P835} provides the speech quality (SIG), background noise quality (BAK), and overall quality (OVRL). Hu and Loizou \cite{hu2006evaluation} showed an accurate linear model of OVRL can be estimated as a function of SIG and BAK. Naderi and Cutler \cite{naderi2021subjective} used this linear relationship to analyze the results of the 3rd Deep Noise Suppression challenge \cite{ch2021interspeech} to estimate the potential improvement in OVRL given a noise suppressor that maximized BAK. Hu and Loizou \cite{hu2006evaluation} released an intrusive speech quality assessment tool based on P.835, with a correlation to subjective quality of  PCC(SIG)=0.70, PCC(CBAK)=0.58, PCC(OVRL)=0.73 using a synthetic training and test set. A commercial tool, 3QUEST \cite{note3quest}, is used to measure the speech (S-MOS), noise (N-MOS), and overall (G-MOS) quality of speech as part of the ETSI EG 202 396-3 standard for mobile telephone quality. This intrusive model has good performance, PCC(S-MOS)=0.92, PCC(N-MOS)=0.94, PCC(G-MOS)=0.94 by condition; it was trained with 179 conditions and tested with 81 conditions, but the duration of training data and testing data is not reported \cite{note3quest}. 

ITU-T Recommendation P.563 is a non-intrusive technique and can directly operate on the degraded signal \cite{p563}. However, it was developed for narrow-band applications, works on limited impairment types, but correlates poorly with human ratings \cite{Avila}. Recently, Deep Neural Networks (DNNs) based approaches have been proposed to estimate the speech quality scores \cite{dong2020attention, gamper2019intrusive, Avila, Reddy2021DNSMOS, ooster2019improving, fu2018quality, catellier2020wawenets, cauchi2019non}. Some of these learning-based approaches use other objective metrics as the ground truth to train their speech quality predictor. Other methods use MOS obtained using P.800 as the ground truth to train their models. In \cite{manocha2020differentiable}, the authors trained the model to identify the Just Noticeable Difference (JND). 
MOS predictors trained on actual human ratings are more reliable than the ones trained to predict other objective metrics like PESQ or POLQA. The accuracy and robustness of the learned models depend on the quality of the human labels and also the quantity and diversity of the audio clips. 
A comparison of some common DNN-based non-intrusive speech quality assessment (NI-SQA) methods is given in Table \ref{tab:comp}. ACR is Absolute Catagorgy Rating \cite{p800}. DNSMOS P.835 is the first P.835 based NI-SQA model we are aware of.

In \cite{Reddy2021DNSMOS}, we show that the NI-SQA metric called DNSMOS trained using subjective quality labels is more robust and reliable than some of the other popular intrusive metrics. DNSMOS is used to do model training and model selection during noise suppression development. DNSMOS is also used for doing ablation studies for noise suppressors \cite{lv2021dccrn+,li2021simultaneous}. DNSMOS has been quite popular, with over a hundred researchers using it after several months of releasing it.

However, DNSMOS only gives the overall score of the audio clip. In this paper, we extend that work to predict the quality of speech (SIG), background noise (BAK), and overall quality (OVRL) of the audio clip. We use the subjective quality labels obtained from ITU-T P.835 from Deep Noise Suppression (DNS) Challenge 3 \cite{ch2021interspeech} and the noisy clips processed by several noise suppression models internally at Microsoft. The labels were obtained using our crowdsourcing-based extension of P.835 described in \cite{naderi2021subjective}. The model uses log power spectrogram as input features to a Convolutional Neural Network (CNN) based model. It can be used to stack rank different DNS methods based on MOS estimates with great accuracy and hence the name DNSMOS P.835. We are providing DNSMOS P.835 as an Azure service for other researchers to use. The details of the API are at \url{www.microsoft.com/en-us/research/dns-challenge/dnsmos}. 

\begin{table}[!b]
  \begin{center}
    \caption{Comparison of some DNN NI-SQA methods}
    \label{tab:comp}
    \begin{tabular}{c|c|c}
      \textbf{Model} & \textbf{Data size} & \textbf{Data}  \\
      & \textbf{(hours)} & \textbf{type} \\
      \hline
      \cite{cauchi2019non} & 5.2 & ACR  \\
      \hline
      WAWENETS \cite{catellier2020wawenets} & 17 & ACR  \\
      \hline
      \cite{Avila} & 27.7 & ACR  \\
      \hline
      \cite{gamper2019intrusive} & 27.7 & ACR  \\
      \hline
      SESQA \cite{serra2021sesqa} & 45.2 & ACR, JND  \\
      \hline
      DNSMOS \cite{Reddy2021DNSMOS} & 300 & ACR \\
      \hline
      DNSMOS P.835 & 75 & P.835 ACR \\
    \end{tabular}
  \end{center}
\end{table}

\section{Data and subjective ratings}
\label{sec:data}

\begin{figure}[t]
 \centering
  \subfloat[Distribution of MOS SIG]{\includegraphics[width=0.24\textwidth]{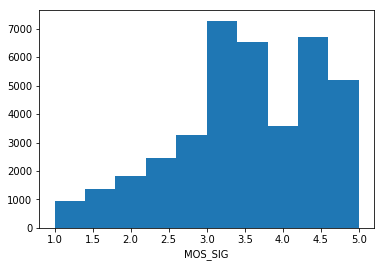}}
  \hfill
  \subfloat[Distribution of MOS BAK]{\includegraphics[width=0.24\textwidth]{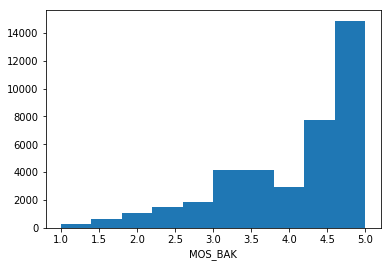}}
  \hfill
  \subfloat[Distribution of MOS OVRL]{\includegraphics[width=0.25\textwidth]{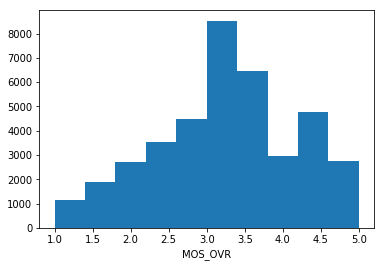}}
  
  \caption{Distribution of the training set}
\label{fig:dist}
\end{figure}

We used the labeled data from the DNS Challenge V3 \cite{ch2021interspeech} to train DNSMOS P.835. The DNS Challenge V3 test set comprised of 600 noisy speech clips processed by about 40 different noise suppression models. The real recordings in the test set were captured in a variety of noise types and Signal to Noise Ratio (SNR) and target levels. The test set is comprised of over 100 noise types and speakers. More details about the creation of these test sets can be found in \cite{ch2021interspeech}. 
The speech quality ratings of the processed clips varied from very poor (MOS=1) to excellent (MOS=5) for SIG, BAK, and OVRL. The distribution of the MOS scores in the training data is shown in Figure \ref{fig:dist}. The scores are highly skewed with most ratings populated in the range 3\textless MOS\textless 4 and fewer ratings in both the tails for SIG and OVRL. However, BAK is highly skewed towards MOS $>$ 4.

The subjective quality ratings are obtained in several P.835 runs conducted over several months. Multiple noise suppression methods are compared in each P.835 run. Each P.835 run included the best-performing noise suppressor, original noisy speech, and a couple of methods with intermediate perceptual quality from previous runs as anchors. Hence, some of the clips were rated multiple times. In total, we have about 30,000 audio clips with associated MOS scores as ground truth. The average length of each audio clip was about 9 seconds, giving us a total of 75 hours of data.

\begin{figure}[hbt!]
 \centering
  \subfloat[Speech MOS]{\includegraphics[width=0.375\textwidth]{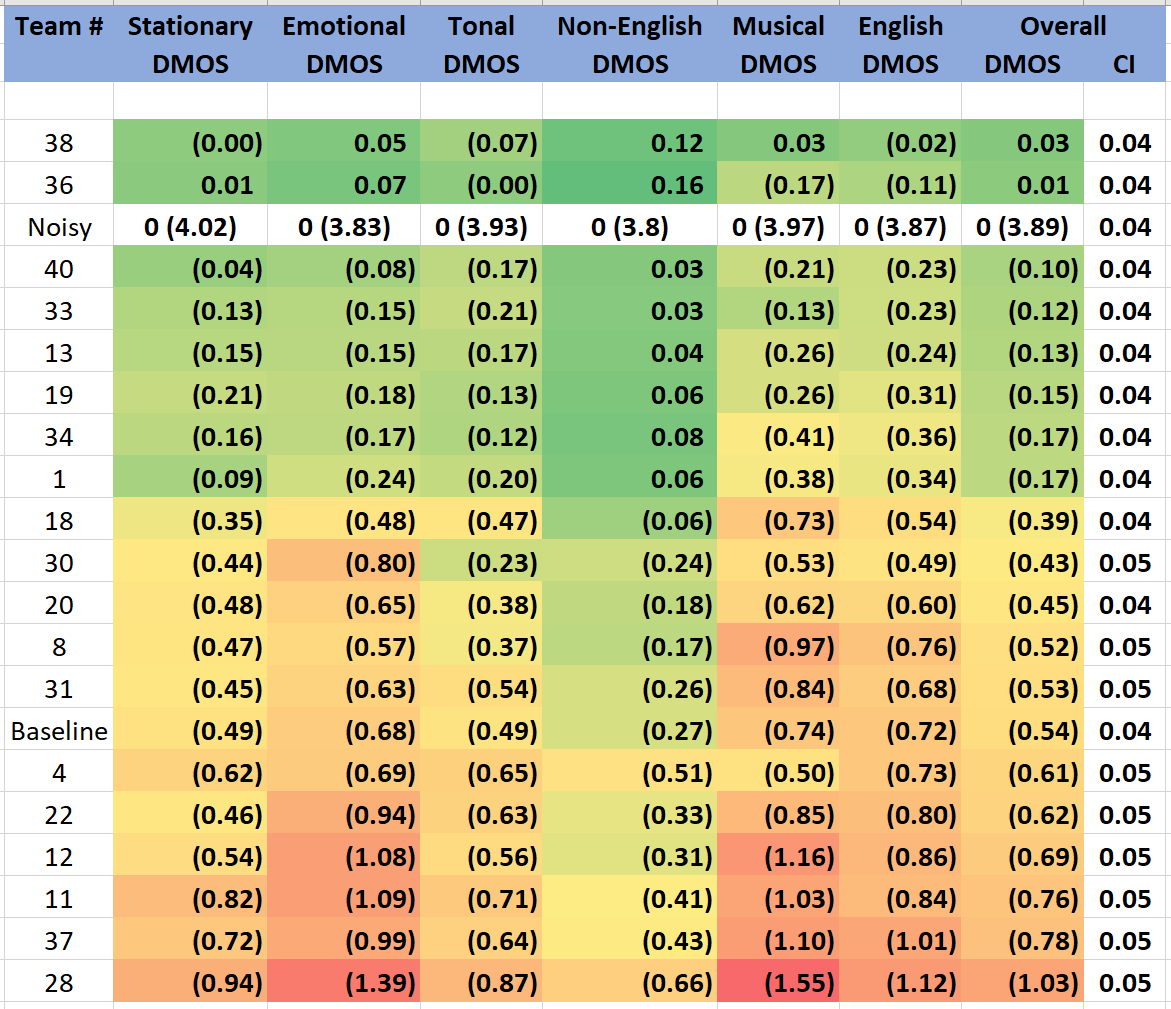}}
  \hfill
  \subfloat[Background Noise MOS]{\includegraphics[width=0.375\textwidth]{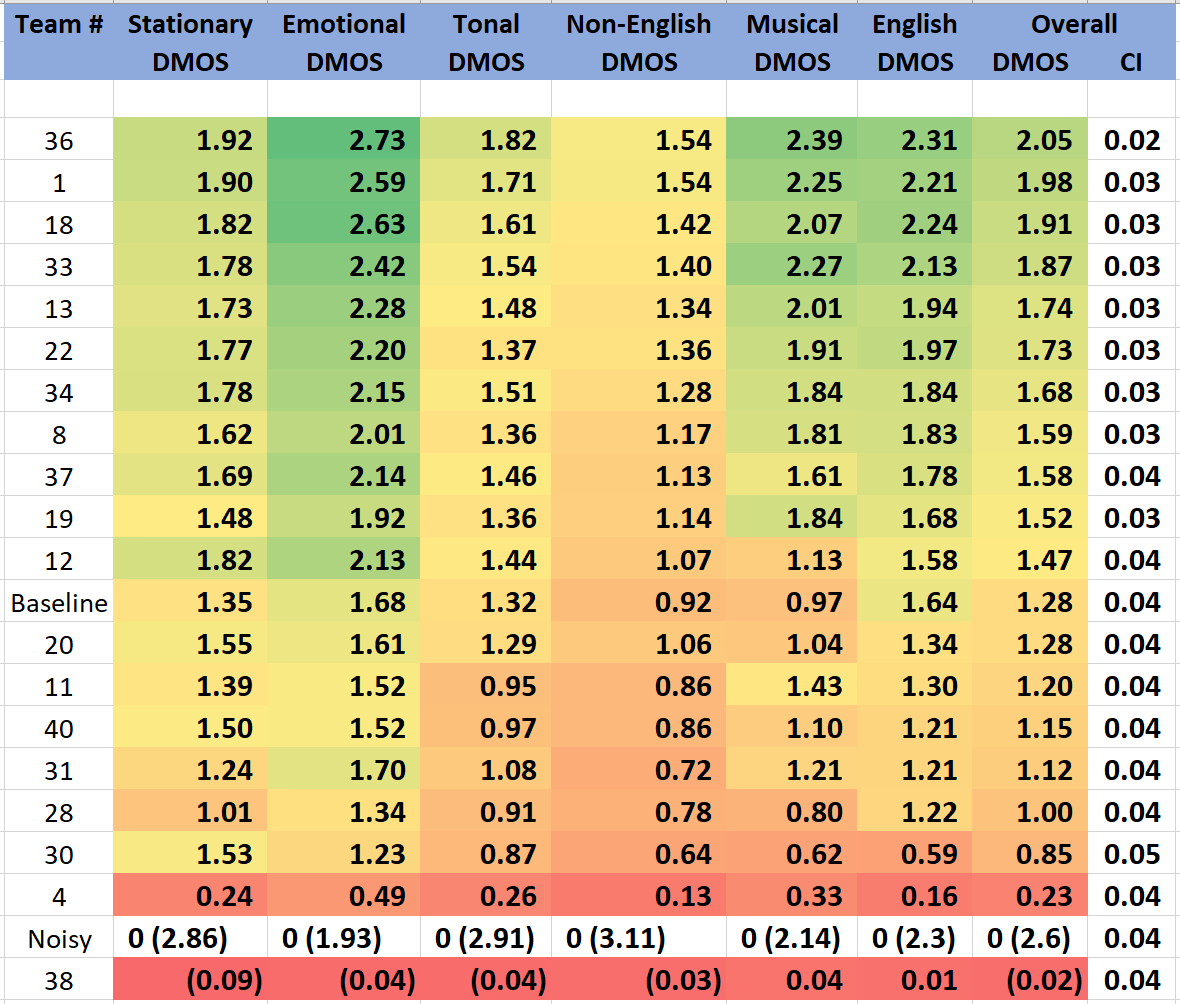}}
  \hfill
  \subfloat[Overall MOS]{\includegraphics[width=0.375\textwidth]{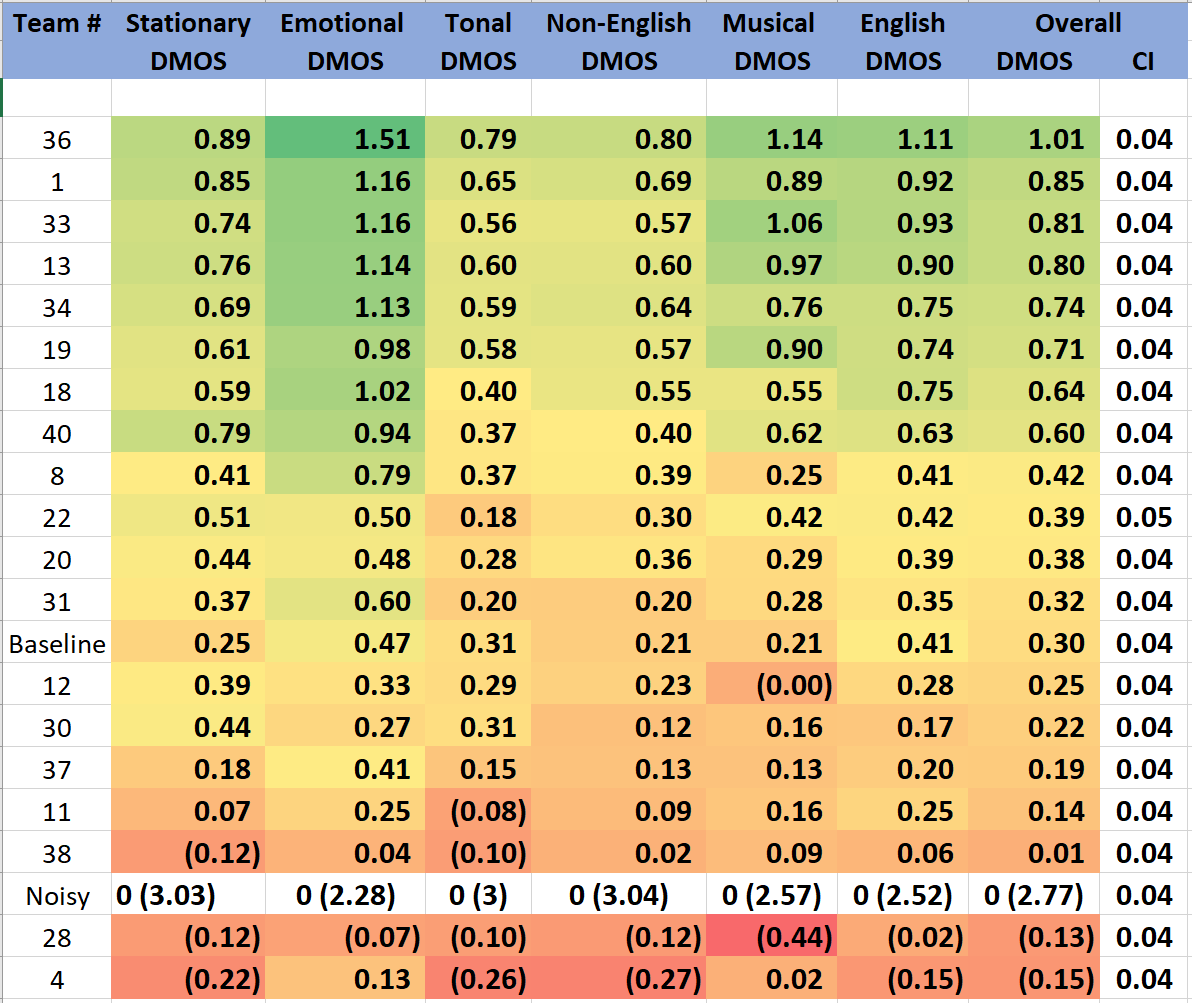}}
  
  \caption{Track 1 results for the 3rd Deep Noise Suppression Challenge}
\label{fig:track1}
\end{figure}

A subset of the dataset is summarized in Figure \ref{fig:track1}. What makes this dataset unique is (1) it is by far the largest P.835 dataset we know of and the only one used to train a DNN non-intrusive speech quality assessment model, and (2) the 40 deep noise suppression models used in the dataset gives a large variety of suppression artifacts we think is needed to generalize a speech quality assessment model for noise suppressors. 

\section{DNSMOS P.835}
\label{sec:model}
\subsection{Features}
\label{ssec:features}
Recently, researchers have seen success in learning features within the model for tasks such as SE \cite{alex2020real}, speech, and music synthesis \cite{tanaka2019wavecyclegan2} and to learn acoustic models \cite{7178847}. They show that using the time-domain waveform requires a larger model trained on a larger and diverse data set to ensure generalization. The ground truth MOS scores are obtained for audio clips with an average length of 9 seconds sampled at 16 kHz. This leads to a very large input dimension if we are treating it as a vector and the model will require many layers to compress and extract input features. Instead, we used log power spectrogram as input feature extracted over 9 seconds duration as it correlates well with human perception and is proven to work very well for analyzing speech quality \cite{gamper2019intrusive}. For spectral features, we used a frame size of 20 ms with a Hamming window and a hop length of 10 ms. The input features are then converted to dB scale. 

\subsection{Prediction model}
\label{ssec:models}
For predicting the MOS scores, we explored different configurations of CNN-based models. The architecture for the best performing model is shown in Table \ref{tab:table1}. The input to the model is log power spectrogram with a 320 FFT size computed over a clip of length 9 seconds sampled at 16 kHz with a frame size of 20 ms and hop length of 10 ms. This results in an input dimension of 900 x 161. We trained two different models with almost the same architecture except for the last layer. One model is trained to predict all 3 outputs (SIG, BAK, OVRL) and the other model is trained to predict only SIG. The reason is we found the prediction of SIG is a much harder task and is less correlated with BAK and OVRL. The models were trained with a batch size of 32 using the Adam optimizer and MSE loss function until the loss saturated. We experimented by adding batch normalization layers after every Conv layer in Table 1. However, adding batch normalization reduces the prediction accuracy of low volume clips. Humans tend to give lower ratings to clips with low amplitudes \cite{nicolas2007influence}. We want the model to capture the variations in the target levels of the data. Hence, we avoid any kind of feature normalization. We also explored different network architectures including CNN followed by LSTM. The model in Table \ref{tab:table1} generalized the best and was of least complexity.    

\begin{table}[t!]
  \begin{center}
    \caption{DNSMOS P.835 Prediction Model}
    \label{tab:table1}
    \begin{tabular}{l|r}
      \textbf{Layer} & \textbf{Output dimension} \\
      \hline
      Input & 900 x 120 x 1\\
      \hline
      Conv: 128, (3 x 3), `ReLU' & 900 x 161 x 128 \\
      Conv: 64, (3 x 3), `ReLU' & 900 x 161 x 64 \\
      Conv: 64, (3 x 3), `ReLU' & 900 x 161 x 64 \\
      Conv: 32, (3 x 3), `ReLU' & 900 x 161 x 32 \\
      MaxPool: (2 x 2), Dropout(0.3) & 450 x 80 x 32 \\
      \hline
      Conv: 32, (3 x 3), `ReLU' & 450 x 80 x 32 \\
      MaxPool: (2 x 2), Dropout(0.3) & 225 x 40 x 32 \\
      \hline
      Conv: 32, (3 x 3), `ReLU' & 112 x 20 x 32 \\
      MaxPool: (2 x 2), Dropout(0.3) & 112 x 15 x 32 \\
      \hline
      Conv: 64, (3 x 3), `ReLU' & 112 x 20 x 64 \\
      GlobalMaxPool & 1 x 64 \\
      \hline
      Dense: 128, `ReLU' & 1 x 128 \\
      Dense: 64, `ReLU' & 1 x 64 \\
      Dense: 1 or 3 & 1 x 1 or 1 x 3 \\
      \hline
      
    \end{tabular}
  \end{center}
\end{table}

\begin{table}[!b]
  \begin{center}
    \caption{Model and clip level correlation of DNSMOS P.835 with human ratings}
    \label{tab:corr}
    \begin{tabular}{c|c|c|c}
      \textbf{Type} & \textbf{SIG} & \textbf{BAK} & \textbf{OVRL} \\
      \hline
      Model PCC &  0.94 & 0.98 & 0.98 \\
      Model SRCC &  0.95 & 0.99 & 0.98\\
      \hline
      Clip PCC &  0.71 & 0.83 & 0.82 \\
      Clip SRCC &  0.72 & 0.82 & 0.81\\
      \hline
    \end{tabular}
  \end{center}
\end{table}

\section{Experimental Results}
\subsection{Test set}
The unseen real test set used to validate the trained model consists of P.835 evaluation of 17 different Microsoft internal noise suppression models on an unseen set of 850 clips. The clips span various categories like emotional, English, Non-English with and without tonal languages, and stationary noises. This unseen test set was created for a future DNS challenge and has similar categories as the training data, adding mouse clicks and improving the quality of emotional speech. The test set was created with crowdsourcing using the method described in \cite{ch2021interspeech}.
\label{sec:testset}
\subsection{Evaluation metric}
\label{ssec:evaluation}
PCC or MSE between the predictions of the developed objective metric and the ground truth human ratings is commonly used to measure the accuracy of the model \cite{gamper2019intrusive, Reddy2021DNSMOS}. From \cite{naderi2021subjective}, we know that P.835 is highly repeatable between runs when averaged across a set of clips per condition, which can be formed by grouping clips enhanced by a particular SE model or based on other criteria like SNR or reverb RT60 times. The PCC computed on the average of ratings per group across different runs is \textgreater 0.9. We also found that PCC computed on the same clips but from two different P.835 runs is only about 0.7-0.8 due to the high rating noise per clip.

Hence, for stack ranking different noise suppressors we evaluate by computing the average of ratings across the entire test set for each model. Therefore, we compute Spearman's Rank Correlation Coefficient (SRCC) and PCC between averaged human ratings and averaged DNSMOS per model. SRCC gives us the stack ranking accuracy of various SE models.

\subsection{Results}
\label{ssec:results}
Table \ref{tab:corr} shows the per model and per clip PCC and SRCC between human ratings and DNSMOS P.835 on the unseen test set described in Section \ref{sec:testset}. When DNSMOS is aggregated by model the results are excellent, though it still shows an area for improvement in SIG. The results on this unseen test set show DNSMOS P.835 generalizes well, at least for these categories of noises and environments. We can not compare DNSMOS P.835 with other metrics since it is the first NI-SQA metric for P.835 we are aware of. 

The clip level correlation of two noise suppression models on the same dataset but using N=30 ratings per clip instead of N=5 used in the per model correlation to give us better accuracy for the human ratings. These results show DNSMOS P.835 has good per clip performance also, though of course not as good as when aggregated at the model level.

\section{Conclusion and Future work}
DNSMOS P.835 is an accurate speech quality metric designed to stack rank noise suppressors with great accuracy. We attribute the excellent performance of DNSMOS P.835 to (1) a large high-quality dataset, (2) a limited speech quality impairment category, (3) significant optimizations on the model architecture and training, and (4) aggregation by noise suppression model. The per clip performance can be improved by significantly increasing the number of ratings per clip, which is currently only 5 because of cost restrictions. We can also expand the complexity of the model to further improve performance.


\bibliographystyle{IEEEbib}
\bibliography{refs}

\end{document}